\documentclass[aps,pre,twocolumn,showpacs,floatfix,superscriptaddress]{revtex4-1}
\usepackage{graphicx}
\usepackage{amssymb}
\usepackage{color}
\usepackage{amsmath}
\usepackage{ulem}
\usepackage{mathrsfs}

\begin{document}

\title{Dynamic scaling of topological ordering in classical systems}

\author{Na Xu}
\affiliation{Department of Physics, Boston University, 590 Commonwealth Avenue, Boston, Massachusetts 02215, USA}
\author{Claudio Castelnovo}
\affiliation{TCM Group, Cavendish Laboratory, University of Cambridge, J.~J.~Thomson Avenue, Cambridge CB3 0HE, United Kingdom}
\author{Roger G. Melko}
\affiliation{Department of Physics and Astronomy, University of Waterloo, Ontario N2L 3G1, Canada }
\affiliation{ Perimeter Institute for Theoretical Physics, Waterloo, Ontario N2L 2Y5, Canada}
\author{Claudio Chamon}
\affiliation{Department of Physics, Boston University, 590 Commonwealth Avenue, Boston, Massachusetts 02215, USA}
\author{Anders W. Sandvik}
\affiliation{Department of Physics, Boston University, 590 Commonwealth Avenue, Boston, Massachusetts 02215, USA}

\begin{abstract} 
We analyze scaling behaviors of simulated annealing carried out on
various classical systems with topological order, obtained as
appropriate limits of the toric code in two and three dimensions. We
first consider the three-dimensional $\mathbb{Z}_2$ (Ising) lattice gauge
model, which exhibits a continuous topological phase transition at
finite temperature. We show that a generalized Kibble-Zurek scaling
ansatz applies to this transition, in spite of the absence of a local
order parameter. We find perimeter-law scaling of the magnitude of a
non-local order parameter (defined using Wilson loops) and a dynamic 
exponent $z=2.70 \pm 0.03$, the latter in good agreement with previous
results for the equilibrium dynamics (autocorrelations). We then study
systems where (topological) order forms only at zero temperature---the 
Ising chain, the two-dimensional $\mathbb{Z}_2$ gauge model, and a
three-dimensional star model (another variant of the $\mathbb{Z}_2$
gauge model). In these systems the correlation length
diverges exponentially, in a way that is non-smooth as a finite-size
system approaches the zero temperature state. We show that the Kibble-Zurek 
theory does not apply in any of these systems. Instead, the dynamics 
can be understood in terms of diffusion and annihilation of
topological defects, which we use to formulate a scaling theory in
good agreement with our simulation results. We also discuss the effect
of open boundaries where defect annihilation competes with a faster
process of evaporation at the surface.
\end{abstract}

\date{\today}


\maketitle


\section{Introduction}

Topological order (TO) cannot be characterized by any local order
parameter and cannot be destroyed through local fluctuations
\cite{wen_all,haldane85,wen90}. Based on these unique characteristics,
systems with topological order have been proposed for use in memory
devices in quantum-information applications
\cite{eric02,nayak08}. Many paradigms for quantum memories and quantum
computing are based on Kitaev's toric code \cite{kitaev06}, which can
be regarded as a quantum generalization of the classical
$\mathbb{Z}_2$ (or Ising) gauge model \cite{wegner71,john79,wansleben85}. 
Whereas most of the focus
to date has been on quantum systems at zero temperature, TO can also
be present in classical systems coupled to a heat
bath~\cite{claudio07,macdonald11,freeman14}.

Here we study the topological ordering dynamics, using protocols
inspired by the Kibble-Zurek (KZ) theory. The KZ mechanism was
originally proposed to describe the formation of defects in the early
expanding universe \cite{kibble76}. Later, it was applied to classical
phase transitions \cite{zurek85,zurek96}, and in recent years it has
been widely used in describing out-of-equilibrium dynamics near
continuous phase transitions in both classical and quantum
systems.~\cite{anatoli05,grandi10,biroli10,jelic11,grandi11,anatoli11,chandran12,hamp15,ricateau17}
The basic idea underlying the KZ mechanism is that a change in some
parameter of a many-body system leads to changes in its relaxation
time $\tau$. Near a critical point $\tau$ has a simple scaling
relationship to the spatial correlation length $\xi$, namely, $\tau
\sim \xi^z$, which defines the exponent $z$ associated with the
dynamics (stochastic or Hamiltonian). By combining this dynamical
scaling with the standard critical form of the correlation length at
distance $\delta$ from a critical point, $\xi \sim \delta^{-\nu}$, it
is possible not only to obtain results for the density of defects, on
which the early studies focused, but also to derive generic scaling
forms for all quantities that exhibit critical scaling in classical
and quantum systems \cite{zhong05,grandi11,chandran12,liu14}. A
central result is that the maximum correlation length a system can
reach in a linear change of a parameter, at velocity $v$, upon
approaching a critical point with correlation-length exponent $\nu$ is
\begin{equation}
\xi_v \sim v^{-1/(z+1/\nu)} 
\, .
\label{eq: xi KZ}
\end{equation}
For a finite system of linear size $L$, this translates into a so-called 
KZ velocity~\cite{chandran12,zhong05,liu14} 
\begin{equation}
v_{KZ} \sim L^{-(z+1/\nu)} 
\, , 
\label{eq: v KZ}
\end{equation}
separating the scaling regimes where the correlation length is velocity 
limited ($\xi_v < L$) and where it is system size limited ($\xi_v > L$). 

Recently, the KZ mechanism has been realized in experiments of cold
atom systems \cite{lamporesi13,clark16}, and proposed to be within
reach of state of the art experiments on spin ice materials~\cite{hamp15}. 
The dynamical scaling functions derived from the KZ mechanism have also found 
applications in simulated annealing (SA) studies of various two-dimensional
(2D) and three-dimensional (3D) systems  with continuous phase 
transitions~\cite{zhong05,biroli10,jelic11,chandran12,liu14,liu15}.
Procedures based on the KZ ansatz have been developed to extract critical 
exponents and critical points~\cite{liu15a}. 
For systems that have continuous phase transition at exactly $T_c=0$, 
such as 2D Ising spin glasses, the KZ ansatz also works, but with a new dynamic 
relaxation exponent that is different from the $T\to 0$ divergent equilibrium 
(autocorrelation) exponent (reflecting non-ergodic Monte Carlo sampling exactly
at $T=0$)~\cite{rubin17,xu17}. However, as far as we are aware, the KZ scaling 
ansatz has never been applied to classical systems that exhibit topological 
phase transitions where there is no local order parameter (in contrast to
the 2D XY model \cite{jelic11}, where the transition is of topological nature 
but there is also a local order parameter). Such transitions 
can take place either at $T>0$ or exactly at $T=0$.

In this paper we demonstrate that KZ scaling applies to finite
temperature topological transitions devoid of a local order
parameter. We study the 3D $\mathbb{Z}_2$ gauge model and determine 
the dynamical exponent to be $z=2.70(3)$, which is consistent with a previous
result based on autocorrelation functions~\cite{ben-av90} but with higher
statistical precision (the number within parentheses above and
henceforth denotes the statistical error---one standard deviation of the 
mean value---in the preceding digit). In contrast, when topological order 
only appears at zero temperature, the conventional KZ mechanism does not 
apply. We are nonetheless able to obtain the dynamical scaling form of the 
non-local order parameter by modeling the relaxation dynamics of 
topological defects. We further investigate the effects of open boundary
conditions, where evaporation of defects at the surface ought to be
taken into account. In all cases, our theoretical arguments are in
good agreement with our extensive numerical SA results.

The paper is organized as follows. In Sec.~\ref{sec: toric code} we briefly 
review the 2D and 3D toric codes and their classical limits; the $\mathbb{Z}_2$ 
gauge models and the so-called 3D star model (another version of the
$\mathbb{Z}_2$ gauge model). In Sec.~\ref{sec:3dtc} we study the
KZ dynamical scaling behavior at the finite temperature transition of
the 3D $\mathbb{Z}_2$ gauge model. In Sec.~\ref{sec:pbc_t0} we study the
models that exhibit only zero-temperature order---the 1D Ising chain, 
the 2D $\mathbb{Z}_2$ gauge model, and the 3D star model---under periodic 
boundary conditions (PBCs). The case of open boundary conditions (OBCs) is 
considered in Sec.~\ref{sec:obc_t0}. Finally, in Sec.~\ref{sec:disc} we 
summarize the main results of this study and discuss their implications. 


\section{Classical limits of the toric code} 
\label{sec: toric code}

The topological classical models studied in this paper are obtained as
appropriate classical limits of the 2D and 3D quantum toric code, which we review 
here for completeness.

The 2D toric code is a system of spin-1/2 degrees of freedom living on the 
bonds of a square lattice with Hamiltonian 
\begin{equation}
H = -J_A\sum_{s} A_{s}-J_B\sum_{p}B_{p} 
\, , 
\label{eq:ham} 
\end{equation}
where 
\begin{align*}
A_{s}= \prod\limits_{i\in s} \sigma^{x} _{i} \, , 
\qquad 
B_{p}= \prod\limits_{j\in p} \sigma^{z} _{j}.  
\end{align*}
\begin{figure}[t]
\centerline{\includegraphics[width=5.5cm, clip]{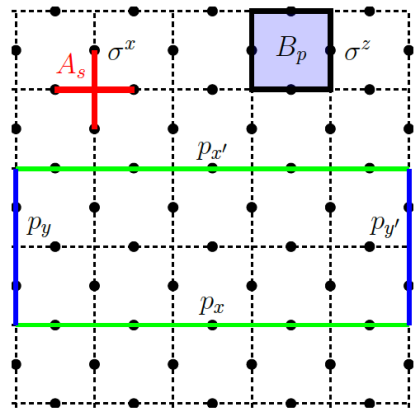}}
\caption{The toric code on a square lattice: The star operator $A_s$,
  shown in red, is the product of $\sigma^x$ components of the spins on the four sites 
  connected to the bonds forming a + (star) centered on site $s$. The
  plaquette operator $B_p$, shown in blue, takes the products of
  $\sigma^z$ components of the spins on the four sites at the edges of a plaquette
  (labeled by $p$). The operators $p_x$ and $p_{x'}$ are defined
  as the product of spins $\sigma^z$ along the green lines. With PBCs,
  the Wilson loop order parameter $\gamma(L)$ is $\langle p_x \,
  p_{x'} \rangle$, where, in our work here, the distance beween the two 
  green lines should be the largest possible in the system. For a 2D square
  lattice of even size, this distance is $L/2$, while for a 3D simple
  cubic lattice it is $\sqrt{2}L/2$. With OBCs, we have to include also the 
  products of boundary spins (along the blue lines), defined as $p_y$ and
  $p_{y'}$, and the order parameter $\gamma(L)$ becomes $\langle p_x
  \, p_{x'} \, p_{y} \, p_{y'} \rangle$. }
\label{fig:fig01}
\end{figure}
$A_{s}$ stands for the star operators, namely the product of $\sigma^x$
components of the spins around the bonds forming a + (star)
at site $s$, and $B_{p}$ denotes the plaquette operators, namely the product of 
$\sigma^z$ components of the spins around the edges of plaquette $p$. These interactions are illustrated in Fig.~\ref{fig:fig01}, where an example of the star operator $A_{s}$ is marked as red and the plaquette operator $B_{p}$ is colored with blue. In 3D, 
the system is defined on a cubic lattice, with similar four-spin plaquette 
operators
but the star operators are upgraded to the product of the six spins on the bonds
stemming from a given site.

All star and plaquette operators commute with one another (and
therefore with the Hamiltonian), and the ground states of the system
have $A_s=+1$ and $B_p=+1$. Excitations above the ground state take
the form of negative stars/plaquettes, with energy penalty $2J_A$ and
$2J_B$ respectively. These defects are referred to as `electric' and
`magnetic', and behave like quasiparticles that can only be created
and annihilated in pairs, under periodic boundary conditions. They are
static under the application of the Hamiltonian, but can otherwise
move freely without energy cost through the action of $\sigma^z$ or
$\sigma^x$ operators (for a review, see for instance
Ref.~\onlinecite{savary17}). In presence of open boundaries, one can
easily see that single defects can nucleate or evaporate at the
surface.

In this paper we shall focus on the following classical limits of the
toric code:
\begin{itemize}
\item In 2D, if one takes either $J_A\to 0$ or $J_B\to 0$, one obtains
  the classical $\mathbb{Z}_2$ lattice gauge
  model~\cite{wegner71,john79}. This model has no finite temperature
  transition, and only orders at $T=0$.

\item In 3D, the limit $J_A\to 0$ yields the classical $\mathbb{Z}_2$
  gauge model~\cite{wegner71,john79}. This model has a
  finite temperature phase transition.

\item In 3D, the limit $J_B\to 0$ yields the version of the $\mathbb{Z}_2$
  gauge model that we here refer to as the 3D star model~\cite{claudio08}. 
	This model has no finite temperature transition but orders topologically at $T=0$.
  
\end{itemize}

The ordered phases in these models are topological in nature, as
reflected, for instance, by a non-zero topological entanglement
entropy~\cite{claudio07,claudio08}. Here we will characterize the 
dynamic topological ordering using the Wilson loops, illustrated
in Fig.~\ref{fig:fig01} for 2D systems. For the 3D star model we will 
use a higher-dimensional generalization of the Wilson loop.


\section{3D $\mathbb{Z}_2$ gauge model at $T=T_c$} 
\label{sec:3dtc}

The 3D $\mathbb{Z}_2$ lattice gauge model exhibits a topological phase
transition at $T_c/J_B = 1.313346$~\cite{john79,wegner71,claudio08} (where we set 
$J_B=1$ hereafter). The transition is in the same universality class as the standard 
3D Ising model, and yet it has no local order parameter in the original spin
degrees of freedom. The mapping between the two models
is a duality between low- and high-temperature partition functions; therefore 
the thermodynamic behavior of the two models is the same, but there is no obvious
relation between their stochastic (Monte Carlo) dynamics. The order parameter for 
the 3D $\mathbb{Z}_2$ lattice gauge
theory is a product of spins across the entire system, namely a
system-spanning Wilson loop. For $T < T_c$, the order parameter decays
exponentially with the perimeter of the contour, $\langle W \rangle
\sim e^{-\alpha L}$, known as the `perimeter law', in contrast to the
`area law' for $T > T_c$, where the order parameter decays
exponentially with the area of the contour, $\langle W \rangle \sim
e^{-\beta L^2}$.~\cite{john79} 

In our simulations, we define a specific Wilson loop as our order parameter: 
\begin{equation}
  \gamma(L) =
  \langle p_x \, p_{x'} \rangle\,, 
\qquad p_x=\prod_{i\in{\cal L}_x} \sigma_i^z
\, , 
\label{eq:gamma}
\end{equation}
where $p_{x}$ and $p_{x'}$ are the products of $\sigma^{z}$ spins
along two lattice lines ${\cal L}_x$ and ${\cal L}_{x'}$ which are
farthest away from each other within the system, as demonstrated in
Fig.~\ref{fig:fig01} for a 2D system. In 3D, the largest possible distance 
is $\sqrt{2}L/2$. Exploiting translation invariance,
  $\gamma(L)$ is averaged over $x$ and $x'$ respecting the maximum
  distance condition.


\subsection{Simulated annealing}
\label{sub:sa}

Here and in the rest of the work we use SA simulations. We first prepare the 
system in equilibrium at a relatively high initial temperature $T_{\rm ini}$
(where a small number of Monte Carlo sweeps is enough to reach equilibrium
when starting from a random configuration), and then we decrease the 
temperature to the final value $T_f$ via the protocol
\begin{equation}
T(t)=T_f+(T_{\rm ini}-T_{\rm f})\,\left(1-t/t_{\rm q}\right)^r
\, , 
\label{sa_prot}
\end{equation}
where $r=1$ stands for the standard SA where temperature decreases
linearly. In general, one can vary the value of $r$ in order to
disentangle the exponents ($z$ and $\nu$) shown in the KZ 
scaling.~\cite{liu14,rubin17} In this study, we only consider the standard 
$r=1$ protocol, since the value of $\nu$ is the same as the one in the
3D Ising model, which is known to high accuracy, 
$\nu=0.62999(5)$~\cite{showk14}. 
We consider $T_{\rm ini}=1.1T_c$ and $T_f=T_c$. The total number of Monte 
Carlo steps during the SA process is denoted by $t_q$, and one step (the unit 
of time) corresponds to a total of $N=L^3$ Metropolis single spin flip 
attempts. The annealing rate (or velocity) $v$ is then defined as 
\begin{equation}
v=(T_{\rm ini}-T_{\rm f})/t_{\rm q}
\, . 
\label{veloticy}
\end{equation}

We simulate systems with sizes $L=8,10,12,16,24$ and $32$. For each
system size, we perform SA runs at various sweeping rates $v$. The range of
velocities varies for different system sizes between about $10^{-6}$ and
$10^{-2}$. We measure the order parameter $\gamma(L)$ as defined in
Eq.~(\ref{eq:gamma}) at various temperatures during each SA process,
averaging over around $10^4$ repeats. Note here that each SA process is independent, with different initial configurations as well as distinct random numbers during the MC updates.

\begin{figure}[t]
\centerline{\includegraphics[width=7.5cm, clip]{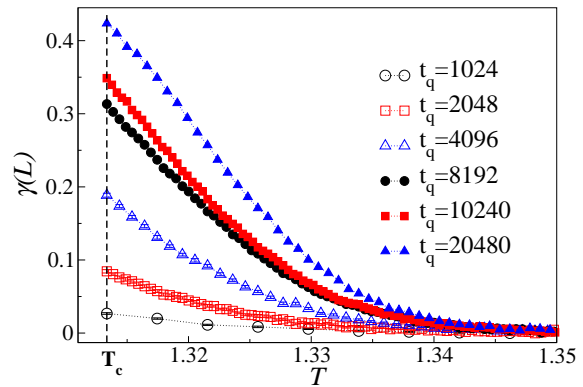}}
\vskip-1mm
\caption{The topological order parameter $\gamma(L)$ for a system with  $N=16^3$ spins
  as a function of the temperature, computed at various total times $t_q$ for annealing 
  from $T=1.1T_c$ to $T_c$. The error bars are smaller than the symbol sizes.}
\label{fig:fig02}
\vskip-1mm
\end{figure}

Figure~\ref{fig:fig02} shows examples of the order parameter $\gamma(L)$ for 
system size $L=16$ at various quenching rates. 
The slower we perform SA, the closer the
system gets to its equilibrium state, i.e., the more ordered it
becomes. The vertical dashed line indicates the last step taken in our SA runs,
ending when $T=T_c$. Since the simplest one-parameter KZ scaling function 
(discussed below) involves only the measurement at $T_c$, in the following we
only focus on the last data point of the SA process at $T=T_c$ for each 
annealing velocity.


\subsection{Dynamic scaling}
\label{sub:ds}

In the generalized KZ non-equilibrium finite-size scaling
form for a physical observable $A$, the dynamic finite-size scaling of $A$ as
a function of annealing velocity is,
\begin{equation}
A(L,v) \sim A_{\rm eq}(L) f(v/v_{KZ})
\, ,
\end{equation}
where $A_{\rm eq}(L)$ denotes the equilibrium finite-size value at $T_c$. 
Normally this value is a power law in the linear size of the system $L$. 
However, we propose that a simple generalization of the KZ form applies 
straightforwardly to other functions of $L$, as relevant to this work. 

For linear SA, the KZ velocity has the form given in Eq.~\eqref{eq: v KZ}, 
$v_{KZ}\sim L^{-z-1/\nu}$. 
Considering the `perimeter law' associated with the Wilson loop order 
parameter $\gamma(L)$ at $T_c$, we expect $\gamma(L)$ measured at 
the critical point to take the form 
\begin{equation}
\gamma(L,v) \sim e^{-\alpha L} f(vL^{z+\frac{1}{\nu}})
\, , 
\label{kz_scaling}
\end{equation}
where $\nu$ is the critical correlation-length exponent and $z$ is the
dynamic critical exponent.

\begin{figure}[t]
\centerline{\includegraphics[width=7.5cm, clip]{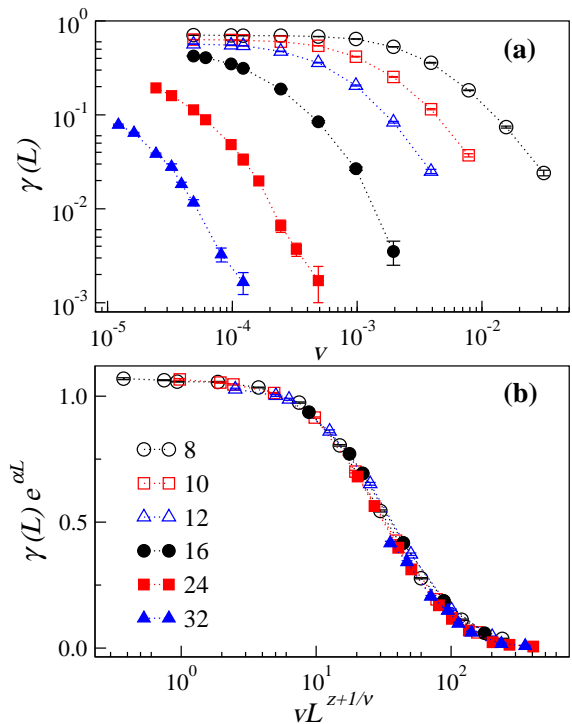}}
\caption{(a) Behavior of the topological order parameter $\gamma(L)$ 
  measured at $T_c$, shown on a log-log plot under various quenching rates 
  for system sizes $L=8$, $10$, $12$, $16$, $24$, and
  $32$ (with the curves decreasing as $L$ increases, as expected for 
	perimeter law behavior). (b) Scaling collapse 
  of $\gamma(L)$ as a function of velocity, based on Eq.~(\ref{kz_scaling}) 
	and shown on a semi-log plot. The optimal value of $z$ for the data 
	collapse is $z\approx 2.70(3)$.} 
\label{fig:fig03}
\end{figure}

Figure~\ref{fig:fig03}(a) shows the behavior of $\gamma(L)$ for
various annealing rates and system sizes from $L=8$ to
$L=32$. Figure~\ref{fig:fig03}(b) shows the velocity scaling of
$\gamma(L)$ based on the KZ scaling function. We vary the values of
exponents $z+1/\nu$ and $\alpha$ to collapse the data according to
Eq.~(\ref{kz_scaling}). The best fit yields the optimal values
$z+1/\nu=4.29(3)$, $\alpha=0.052(1)$. As $\nu\approx0.63$, we obtain
$z=2.70(3)$. The statistcal errors were determined by a bootstrap analysis.
For further details on the data-collapse procedures we refer to
Refs.~\onlinecite{liu14} and \onlinecite{rubin17}.

Previous Monte Carlo studies of the equilibrium relaxation 
(autocorrelation) time at $T_c$ gave $z=2.5(2)$ \cite{ben-av90}. Thus,
our result for $z$ agrees with the previous value within error bars, 
but we improve the statistical precision by one digit. The general 
expectation is that the dynamic exponent appearing within the 
out-of-equilibrium KZ framework should indeed be the same as the one 
at equilibrium when $T_c > 0$ (while for systems with $T_c=0$ 
this is not the case~\cite{rubin17,xu17}). 
The good collapse of the data reveals 
that, as with other continuous phase transitions described by local order
parameters \cite{biroli10,chandran12,liu14,liu15}, KZ scaling also works for 
topological phase transitions devoid of a local order parameter. We stress
again that the standard KZ scaling form in this case is also modified by the 
exponential form of the equilibrium size-dependence in Eq.~(\ref{kz_scaling}).

Recall that the mapping between the 3D $\mathbb{Z}_2$ lattice gauge
model and the 3D Ising model is a duality between the partition
functions, and thus has no dynamical implications. Moreover, the dynamic exponent is not an intrinsic property of a model, as it also depends on the specific update algorithm. While they share
the same thermodynamic critical properties, it is not
surprising that they have different dynamical exponents, $z\approx
2.7$ and $z\approx 2.0$~\cite{wansleben87,wang95}, for the gauge model
and standard 3D Ising model, respectively. There may exist an update algorithm for the 3D $\mathbb{Z}_2$ lattice gauge
model that matches exactly with the local update of the 3D Ising model. However, as the duality mapping between the two models is highly nontrivial, we expect the algorithm to be highly nontrivial as well.

Note also that the out-of-equilibrium
SA approach with KZ scaling circumvents the need to ensure that the system 
is in equilibrium when using autocorrelation functions to estimate the 
equilibrium dynamic exponent. Each repetition of the SA procedure 
represents a statistically independent contribution to the estimated
mean values. Thus, the only potential source of systematic errors 
is corrections to scaling in the analysis. Based on the good data
collapse for large systems at the known value of $T_c$, we judge that
the impact of scaling corrections should be small in the above results
for the 3D $\mathbb{Z}_2$ lattice gauge model.


\section{\label{sec:pbc_t0} $T=0$ topological order with periodic boundary conditions}
In this section we study models that have no finite-temperature phase
transition and topological order only forms at $T=0$, when the defect
density vanishes identically at equilibrium. Namely, we consider the
3D star model~\cite{claudio08} and the 2D $\mathbb{Z}_2$ lattice
gauge model. In addition, we also consider their natural reduction down
to 1D; the standard ferromagnetic Ising chain. 


\subsection{Failure of the Kibble-Zurek mechanism}

For systems that order only at $T=0$, we cannot apply directly the standard 
KZ scaling forms, because when $T\rightarrow 0$ the correlation length 
diverges exponentially, $\xi \sim \exp(c/T)$, instead of following the 
power-law behavior expected at finite-$T$ continuous phase transitions. 
In principle the exponential form is not an issue in itself, as 
apparent in the detailed derivation of the KZ scaling forms in 
Ref.~\onlinecite{liu14} (see also Ref.~\onlinecite{chandran12}). 
As long as there is a known relationship between 
the correlation length and the relaxation time, a criterion for quasi-static 
equilibrium---giving a critical velocity separating slow and fast processes, 
equivalent to Eq.~\eqref{eq: v KZ}---can be obtained. 
For example, in the 1D Ising model the correlation length has exactly the form 
$\xi \sim \exp(c/T)$. If one assumes that the relaxation time is a power of 
this length, $\tau \sim \xi^z$, as expected with $z=2$ based on the fact 
that the domain walls perform 1D random walks, one finds that the 
critical KZ velocity is $v_{\rm crit} \sim L^{-z}\ln^{-2}(L)$. 

However, this result is incorrect, differing by a factor of $\ln(L)$ from 
the known rigorous expression obtained by Krapivsky for this 
model~\cite{paul10}. 
The reason for the failure of this simplistic
approach is that the correlation length is not changing smoothly in a given
realization of the annealing process in a finite system at the last stages
of equilibration. When the number of domain walls (defects) is small, 
the (kink-antikink) annihilation of a defect pair leads to large jumps in 
the correlation length. 
For instance, the very last annihilation process in a 1D Ising model of 
finite size $L$ produces a jump in the correlation length from $\xi=L/2$ 
to $\xi=L$. 
On the contrary, a continuous (in the large $L$ limit) growth of the 
correlation length all the way to $\xi=L$ is a key assumption in the 
derivation of the KZ scaling expressions~\cite{liu14}.


\subsection{Scaling theory for defect annihilation}

We are nonetheless able to obtain a finite-size scaling form for the 
order parameter in these systems, as they are ramped down to zero temperature, 
by looking more closely at the nature of their defects and how order 
emerges as the defect density vanishes. 
As in the 1D Ising model, the excitations at low $T$ in the 2D 
$\mathbb{Z}_2$ gauge model and the 3D star model also take the form
of stochastically itinerant non-interacting point-like quasiparticles. 
The point-like nature of the excitations is closely related to the
absence of a phase transition. Indeed, the energy-free (diffusive)
motion of these quasiparticles is able to change the value of
the (topological) order parameter. Therefore, whenever
excitations are present in the system, the order parameter remains
vanishingly small. This is clearly the case at all $T>0$ in the
thermodynamic limit. A non-vanishing order parameter can, however,
appear as a finite-size effect when the temperature becomes so low
that on average less than one pair of defects is left in the system.
This behavior is controlled by the very final stage of relaxation into 
the topologically ordered state, namely the disappearance of the last 
excitations. 
With periodic boundary conditions, this corresponds to the process where 
the last pair of defects meet and annihilate.

Considering SA with linear sweeps down to $T_f=0$, to quantify the longest
time scale we can assume that the system remains in equilibrium (with vanishingly 
small order parameter) down to a threshold temperature $T_{\rm th}$ where the 
number of defects left in the (finite) system is of order $1$,
\begin{equation}
\exp\left( -\frac{\Delta}{T_{\rm th}} \right) \sim L^{-d}.
\label{eq: Tth}
\end{equation}
Here $\Delta$ is the bare cost of a defect (e.g., the cost of a single
domain wall in the 1D Ising model), and $d = 1,2,3$ is the dimensionality 
of the system. Only if the sweep continues for a sufficiently long
time from $T_{\rm th}$ down to $T=0$ is the order parameter finally able 
to acquire a finite expectation value via the annihilation of the last two 
remaining defects.  Therefore, the scaling behavior of the order parameter 
at the end of the sweep ($T=0$) is controlled by this regime.

Taking $T_{\rm f}=0$ and $r=1$ in Eq.~(\ref{sa_prot}), the time
dependence of the temperature in a SA sweep in $t \in (0,t_{\rm q})$
takes the form
\begin{equation}
T(t) = T_{\rm ini} \left( 1 - \frac{t}{t_{\rm q}}\right)
, 
\label{eq: sweep parameter}
\end{equation}
where $T_{\rm ini}$ is the initial temperature and $t_q$ is the
number of total quench steps. 
The sweep velocity is thus $v = T_{\rm ini} / t_{\rm q}$ and 
the time it takes from $T_{\rm th}$ to $T=0$ is
\begin{equation}
\Delta t=t_{q}-t_{\rm th} = \frac{T_{\rm th}}{v} 
\, . 
\label{eq:dt}
\end{equation}
Inserting the expression for $T_{\rm th}$ from Eq.~\eqref{eq: Tth} 
into the above expression, we get
\begin{equation}
\Delta t \sim \frac{\Delta}{v\ln(L)} 
\, ,
\label{eq:dt2}
\end{equation}
which we can now relate to the time scale of defect annihilation.
The system develops a non-vanishing order parameter in the span of
time $\Delta t$ only if the last quasiparticles in the system meet and
annihilate. 

As the quasiparticles are non-interacting, their motion is
diffusive and the time scale for annihilation $\tau_{\rm annihilation}$ 
should depend on dimensionality and system size~\cite{condamin05}: 
\begin{equation} \label{eq:dt_L}
   \tau_{\rm annihilation} \sim
               \begin{cases}
               L^2,~~~  & d=1
              \\
               L^2\ln (L),~~~& d=2
              \\
              L^3,~~~  & d=3 \, . 
            \end{cases}
\end{equation}
\begin{figure}[t]
\centerline{\includegraphics[width=6.5cm, clip]{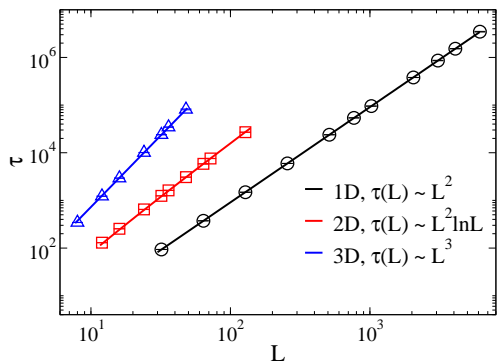}}
\caption{Behavior of the mean time $\tau$ required for annihilation of the 
  last pair of defects through random walks vs the system size $L$ in 1D (black), 2D (red) and
  3D (blue) lattices with PBCs. The solid curves are fits
  based on the expected scaling forms in Eq.~(\ref{eq:dt_L}).}
\label{fig:fig04}
\end{figure}
We numerically tested these scaling laws by considering the case of
two defects (with random initial conditions) performing random walks
on 1D, 2D and 3D lattices with PBCs. We measured the average relaxation
time $\tau$, which is the number of total steps the defects take
before they meet and annihilate (one step corresponding to one lattice
move of each defect). Our results are presented in Fig.~\ref{fig:fig04}. 
The excellent agreement with the scaling form in Eq.~\eqref{eq:dt_L} 
demonstrates the lack of significant finite-size corrections even for the 
smallest system sizes considered in this work---an important benchmark for 
the interpretation of our results on topological systems below. 

The probability that the system develops a non-vanishing order
parameter in an SA run is controlled by the ratio $\Delta
t/\tau_{\rm annihilation}$. Combining Eqs.~(\ref{eq:dt2}) and
(\ref{eq:dt_L}), this ratio can be expressed in a KZ-like scaling form
as
\begin{equation} 
  \frac{\tau_{\rm annihilation}}{\Delta t}
  =
  \frac{v}{v_{\rm crit}}
  \;,
\end{equation}
where
\begin{equation} \label{eq:v_L}
  v_{\rm crit} \sim
              \begin{cases}
               L^{-2}\ln^{-1}(L),~~~ & d=1,
              \\
               L^{-2}\ln^{-2}(L),~~~ & d=2,
              \\
              L^{-3}\ln^{-1}(L),~~~ & d=3 \, . 
              \end{cases}
\end{equation}
We thus expect that the dynamic finite-size scaling function of an
appropriate order parameter $M$ in each of the systems considered here 
takes the form 
\begin{equation}
M\sim f(v/v_{\rm crit}) 
\, ,
\label{eq:t0_scaling}
\end{equation}
which is formally similar to the KZ scaling ansatz but with critical 
velocities that cannot be derived within that formalism. 

We note that the case of the 1D Ising chain was previously studied
analytically in a somewhat different way in Ref.~\onlinecite{paul10}, 
and the domain wall density there indeed shows a scaling form consistent with 
our Eqs.~(\ref{eq:v_L}) and (\ref{eq:t0_scaling}). 
Another study related to our work is Ref.~\onlinecite{freeman14}, where the 
finite-size scaling of the 2D toric code was considered 
using an effective classical model in contact with a thermal reservoir. 
There the focus was on the time scale on which topological order is destroyed 
at fixed temperature through topological point defects undergoing nontrivial 
random walks; this is different from the case studied here 
where we consider the opposite process of topological ordering under SA 
down to $T=0$. The time scales in our work and in Ref.~\onlinecite{freeman14} 
are therefore not the same. 


\subsection{Simulated annealing results}

We performed SA runs (setting $T_{\rm ini}=2$) 
with various annealing velocities for the 1D Ising model,
the 2D $\mathbb{Z}_2$ lattice gauge model and the 3D star model, using several
system lengths $L$ in each case. For the Ising chain, we choose the 
commonly-used squared magnetization, $m^2$, as our order parameter, 
\begin{equation}
m^2= \left\langle \frac{1}{L}\sum_{i=1}^{L}\sigma_i \right\rangle^2 
\, . 
\label{eq:1dop}
\end{equation}
For the 2D $\mathbb{Z}_2$ gauge model, we use a Wilson loop
order parameter similar to that introduced for the 3D case in
Sec.~\ref{sec:3dtc} and illustrated in Fig.~\ref{fig:fig01}.
The only difference from the 3D case 
is that now the farthest distance between the lines ${\cal L}_x$ and 
${\cal L}_{x'}$ is $L/2$ instead of $\sqrt{2}L/2$. 

\begin{figure}[ht!]
\centerline{\includegraphics[width=5.2cm, clip]{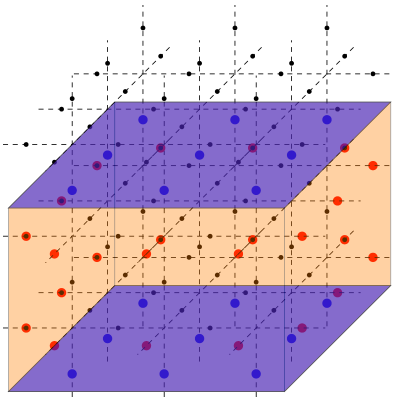}}
\caption{Illustration of the topological order parameter of the 3D star model
  with periodic and open boundary conditions. For PBCs the order
  parameter $\pi(L)$ is the average of surface-surface correlations,
  indicated in blue. For OBCs the order parameter should include also
  the product of spins on the boundaries between these two surfaces,
  shown in orange, so as to form a closed surface.}
\label{fig:fig05}
\end{figure}

For the 3D star model, the topological state has a different nature with 
respect to a $\mathbb{Z}_2$ gauge model, and the role of Wilson loops is 
played by products of spins around (dual) closed surfaces that are locally 
perpendicular to and bisect the bonds of the original lattice. 
The simplest such surface is a unit dual cube surrounding a single vertex 
on the original lattice, and the six spins on the bonds stemming from that 
vertex live respectively at the centers of the six faces of the cube. 
In the ground state, the product of the six spins is $1$ (namely, the product 
of the six spins on the faces of the dual cubic surface). 
For a detailed discussion of these topological structures we refer the 
reader to Ref.~\onlinecite{claudio08}. 
Here we follow that reference and introduce the corresponding order parameter 
as the product of all the spins on two parallel (dual) lattice planes, 
${\cal P}_x$ and ${\cal P}_{x'}$, at, say, fixed $x$ and $x'$ values on 
the lattice (see Fig.~\ref{fig:fig05}). 
For a system with periodic boundary conditions, the product of the spins on 
the two planes equals the product of all dual unit cubes around the vertices 
in between the two planes. Therefore, in the ground state the product 
takes value $1$. 
This product acts as a topological order parameter, similar to the Wilson 
loop used for the Ising gauge models with plaquette interactions.

For convenience, we denote as $s_x$ and $s_{x'}$ the products of 
all the spins on each of the two planes ${\cal P}_x$ and ${\cal P}_{x'}$, 
separately, and we define 
\begin{equation}
\pi(L) = \langle s_x \, s_{x'} \rangle 
\, , 
\qquad s_x=\prod_{i\in {\cal P}_x} \sigma_i^z
\, ,
\label{eq:3dop}
\end{equation}
as a closed-surface analog of the Wilson loop. 
Here the distance between the two surfaces ${\cal P}_x$ and 
${\cal P}_{x'}$ is taken to be maximal, namely $L/2$. 

\begin{figure}[t]
\centerline{\includegraphics[width=7.0cm, clip]{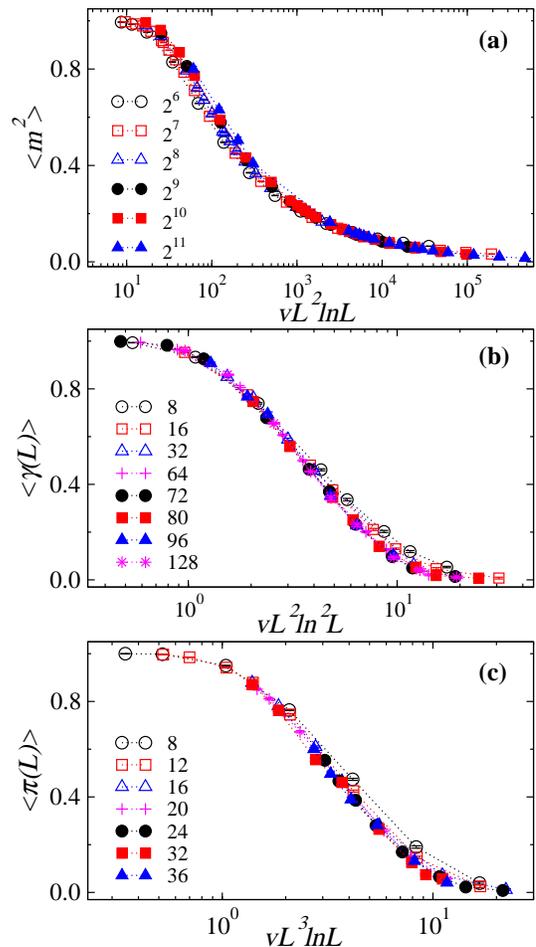}}
\vskip-1mm
\caption{Scaling behavior on semi-log plots of the order parameters in (a)
  the 1D Ising chain, (b) the 2D $\mathbb{Z}_2$ gauge model,
  and (c) the 3D star model with PBCs.}
\label{fig:fig06}
\vskip-1mm
\end{figure}

The behavior of the three order parameters after rescaling according
to Eqs.~(\ref{eq:v_L}) and (\ref{eq:t0_scaling}) is presented in
Fig.~\ref{fig:fig06}.
We find good scaling collapse of the data, and the trend is a clear improvement
with increasing system size, suggesting that the scaling functions we propose 
are indeed correct. In principle, a scale factor $L_0$ inside the logarithms
of the scaling arguments could also be included, $\ln(L/L_0)$, but we
find that the optimum value of this factor is close to $1$ and the
data collapse is not significantly improved. We therefore did not
include such a scale factor in the figure and further below.

One can notice that the data collapse gets worse when $v \gg v_{\rm crit}$, 
which is expected as the scaling form was derived under the assumption that 
the system remains at equilibrium down to $T_{\rm th}$. This assumption breaks down 
at high velocity, in such a way that the scaled data peel off from the common 
scaling form at a point that moves to the right as the system size increases. This 
is similar to what happens in KZ scaling, as discussed in Ref.~\onlinecite{liu14}. 
We conclude that the low-$T$ dynamics of these systems is indeed controlled 
by defect-defect annihilation processes of free random walking quasiparticles. 


\section{\label{sec:obc_t0}$T=0$ topological order with open boundary conditions}

In this section, we discuss how OBCs affect the dynamics of the
topological order parameters for the systems studied in
Sec.~\ref{sec:pbc_t0}. With PBCs, the only way for defects to vanish is
through defect pair annihilation. With OBCs, however, defects can
diffuse to and disappear through the open boundaries---thus single
defects can ``evaporate''.  

\subsection{Scaling of boundary evaporation}

Whereas the time for pair annihilation
scales as $L^2$, $L^2\ln L$, and $L^3$ in $d=1$, $2$, and $3$ [see
  Eq.~(\ref{eq:dt_L})], defects can reach the boundary within a
typical time scale
\begin{equation}
\tau_{\rm boundary} \sim L^2 
\, , 
\label{dt_opc}
\end{equation}
irrespective of dimensionality \cite{sidbook}. Clearly, when comparing
the two kinds of dynamics, boundary evaporation takes either equal
(1D) or shorter (2D and 3D) time. Therefore, the low-temperature
dynamics should be dominated by boundary processes, leading to a
different critical velocity in the dynamic scaling function
$f(v/v_{\rm crit})$. Following the discussion in the previous section,
upon replacing $\tau_{\rm annihilation}$ by $\tau_{\rm boundary}$, we
obtain a form for $v_{\rm crit}$ that is universal for 1D, 2D and 3D
lattices with OBCs:
\begin{equation}
v_{\rm crit}\sim L^{-2}\ln^{-1}(L)
\, . 
\end{equation} 
 
Notice that the order parameters for the 2D $\mathbb{Z}_2$ lattice
gauge model $\gamma(L)$ and the 3D star model $\pi(L)$ have to be
redefined after switching to OBCs (while for the 1D Ising chain it
remains the same). For the 2D $\mathbb{Z}_2$ gauge model, as
illustrated in Fig.~\ref{fig:fig01}, in addition to the two line
operators $p_x$ and $p_{x'}$, we also need to include the spins on the
boundaries between the two lines, i.e., $p_y$ and $p_{y'}$, in order
to form a closed loop. Therefore, the order parameter becomes,
\begin{equation}
\gamma(L)=\langle p_x\, p_{x'} \, p_y\, p_{y'}  \rangle 
\, . 
\label{eq:gamma2}
\end{equation}
For the 3D star model, as illustrated in Fig.~\ref{fig:fig05}, in
addition to the two surface operators $s_x$ and $s_{x'}$, we need to
include the product of spins on the boundary surfaces in between,
to form a closed surface.
\begin{figure}[t]
\centerline{\includegraphics[width=7.0cm, clip]{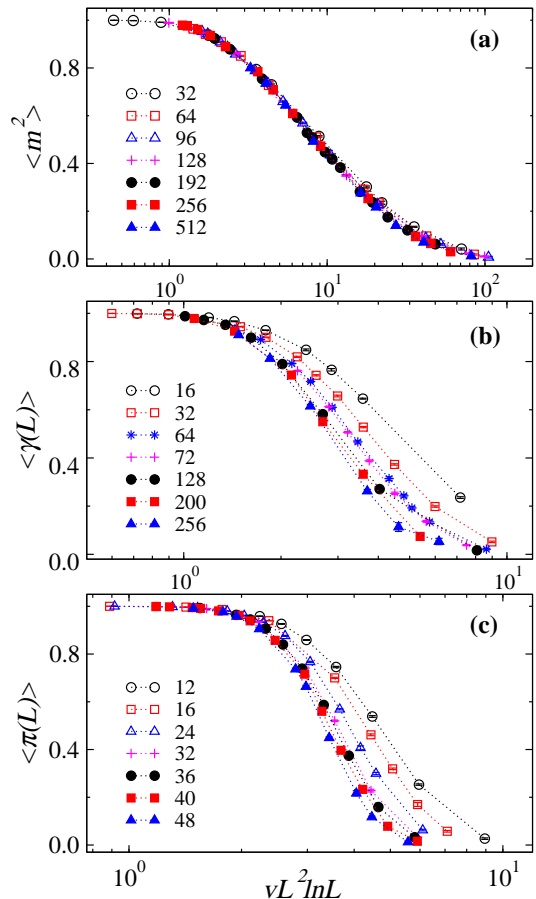}}
\caption{Scaling of the order parameters for OBC systems for (a) the 1D
  Ising model, (b) the 2D $\mathbb{Z}_2$ gauge model, and (c) the 3D
  star model. In all cases we have used the universal scaling form $f(v \,
  L^2\ln L)$.}
\label{fig:fig07}
\end{figure}

Figure~\ref{fig:fig07} shows the scaling behavior of the order
parameters with OBCs. The data are plotted according to the new
universal scaling form $f(v/v_{\rm crit})$, where $v/v_{\rm crit} \sim
v \, L^2\ln L$, due to the (faster) dynamical process whereby defects
reach the open boundaries by random walking and evaporate. The
excellent collapse in panel (a) is expected due to the fact that the
two relaxation processses lead to the same scaling form in 1D. The
scaling collapse in panels (b) and (c), 2D and 3D respectively, is far
less satisfactory. The data points for small system sizes show a
substantial deviation from the predicted behavior. As system size
increases, the data collapse gradually improves, suggesting that the
scaling form is correct in the thermodynamic limit but finite-size
effects are far stronger with OBCs than PBCs. This is likely due to the
contributions from the two dynamical processes with time scales that
diverge from one another for $L \gg 1$, but are in fact quite close
for small systems (indeed 
$\tau_{\rm annihilation} / \tau_{\rm boundary} = \ln (L)$ in 2D, which is 
just $\simeq 4.6$ for $L=100$;
and $\tau_{\rm annihilation} / \tau_{\rm boundary} = L$ in 3D, where
we can only access relatively small system sizes overall).

\subsection{Combining evaporation and annihilation}

\begin{figure}[t]
\centerline{\includegraphics[width=7.0cm, clip]{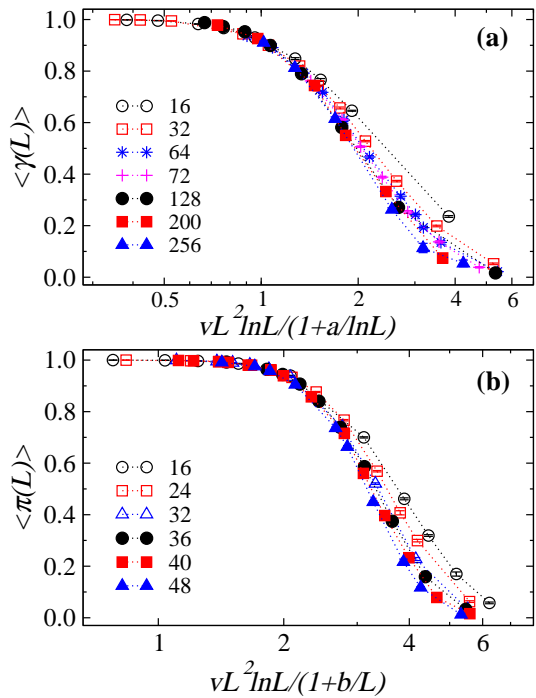}}
\caption{Modified velocity scaling of the order parameter for the 2D and 3D OBC
systems, combining effectively both relaxation processes (pair annihilation 
and boundary/surface evaporation). (a) Results for the 2D $\mathbb{Z}_2$ lattice gauge 
model, with the velocity scaled as $ v\,L^2\ln(L)(1+a/\ln(L))^{-1}$ with $a$=2.4(2), and 
(b) for the  3D star model with rescaling $ \sim v\,L^2\ln(L)(1+b/L)^{-1}$ with $b=2.2(3)$.
In (a) the smallest system was excluded from the data-collapse analysis, while in (b) all
sizes were used.}
\label{fig:fig08}
\end{figure}

In order to account for the two different defect removal processes, we
consider a modified scaling approach for the 2D and 3D systems. The
rate of defect depletion is the sum of those for the two separate
channels (defect annihilation and evaporation at the boundary),
yielding the effective rate
\begin{equation}
  \tau^{-1}_{\rm eff} =
  \tau^{-1}_{\rm boundary}
  +
  \tau^{-1}_{\rm annihilation}
  \;.
\label{taurate}
\end{equation}
From this total rate, we obtain the effective critical velocity scale,
which is the sum of the critical velocities for the two separate
processes.

Recall that, for the 2D case, the critical velocities for the process
of defect annihilation and boundary evaporation are given by $v_{\rm
  crit,a} \sim L^{-2}\ln^{-2}(L)$ and $v_{\rm crit,e} \sim
L^{-2}\ln^{-1}(L)$, respectively. We therefore propose a combined
critical velocity of the form
\begin{align}
v_{\rm crit} &\sim L^{-2}\ln^{-1}(L)+aL^{-2}\ln^{-2}(L)  \nonumber \\
                        &\sim L^{-2}\ln^{-1}(L)\left (1+\frac{a}{\ln(L)}\right ),
\end{align}
where $a$ is a fitting parameter that accounts for ${\cal O}(1)$
prefactors in the $L$ dependence of the velocities. The modified argument of the 
scaling function in 2D should then take the form
\begin{equation}
(v/v_{\rm crit})_{\rm 2D} \sim v\,L^2\ln(L)\left(1+\frac{a}{\ln(L)}\right)^{-1}.
\end{equation}
Given that the two processes contribute additively to the overall rate in 
Eq.~(\ref{taurate}) we expect $a>0$. 

Analogously, for the 3D star model, the improved scaling argument should 
take the form
\begin{equation}
(v/v_{\rm crit})_{\rm 3D} \sim v\,L^2\ln(L)\left(1+\frac{b}{L}\right)^{-1},
\end{equation}
where $b$ again is a fitting parameter that is expected to be
positive. Fig.~\ref{fig:fig08} shows the analysis for both cases
based on the new scaling functions, which indeed improve the data
collapse considerably (especially in the 2D case). The best data
collapse yields $a=2.4(2)$ and $b=2.2(3)$, both positive as expected.


\section{\label{sec:disc} Discussion}

In this paper we studied the dynamical scaling behavior of
several classical topological systems using SA with varying sweeping
rates. For the 3D $\mathbb{Z}_2$ lattice gauge model, which undergoes a
topological phase transition at finite temperature, the dynamical
scaling of its order parameter can be understood as a simple generalization
of the KZ mechanism to an order parameter with exponential size dependence
at $T_c$ (instead of the power law applying at standard continuous phase
transitions). To our knowledge, this is the first time that KZ theory has 
been applied and tested numerically on a classical system with a continuous
topological phase transition devoid of a local order parameter. Our result
for the dynamical exponent of the system, $z=2.70(3)$, is consistent
within error bars (which are an order of magnitude smaller in our work)
with a previously published value based on equilibrium autocorrelations
\cite{ben-av90}, thus supporting the notion of a common exponent describing 
the equilibrium and the out-of-equilibrium relaxation. 

In contrast to systems with continuous phase transitions with $T_c>0$, we 
point out that finite-size scaling functions based on the generalized KZ 
ansatz do not apply to transitions into topological phases that only exist 
at zero temperature. 
In these systems the correlation length diverges exponentially in temperature 
as $T \to 0$, and at the last stage of ordering the finite-size correlation 
length jumps when topological defects (the end-points of strings) 
finally disappear. We studied examples of such systems, namely, the 2D $\mathbb{Z}_2$ 
lattice gauge model and the 3D star model. For completeness we also
studied a simpler but analogous system in 1D: the standard Ising chain that 
was also previously investigated by Krapivsky \cite{paul10}. 
The stochastic dynamics of ordering in these systems is dominated by the 
diffusion of the end-points of open strings, and we 
proposed scaling functions that are obtained from dynamical modeling of the 
annihilation processes of these topological point defects through random 
walks. We find excellent agreement between the proposed 
scaling laws and numerical SA simulations, suggesting that we have correctly
identified and modelled the relevant relaxation processes in these systems. 
Note also that while the numerical analysis cannot strictly distinguish between
the KZ and the defect-annihilation forms, because they differ only
logarithmically, our physical arguments against the standard KZ
scaling mechanism are unambiguous.

We also studied the effect of open boundaries, where individual defects can 
evaporate. We find that defect evaporation dominates over pair annihilation 
for large enough systems. For system sizes that are numerically 
accessible, a scaling approach combining the different time-scales of the 
two processes is needed to fit the data. 

It is important to contrast our results for the $T=0$-ordering models 
with the KZ scenario. A naive application of the KZ mechanism according to 
the derivation in Appendix A of Ref.~\onlinecite{liu14} gives 
$v_{\rm crit}\sim L^{-2}\ln^{-2}(L)$ if we assume forms 
of the correlation length and relaxation time scale appropriate for the 
$T_c=0$ systems considered here: $\xi \sim \exp(c/T)$ and $\tau \sim \xi^z$ with 
$z=2$. Interestingly, comparing with the forms we have obtained based on the defect 
annihilation scenario, Eq.~(\ref{eq:v_L}), the results are exactly the same 
in 2D, while they differ by a factor $\ln(L)$ and $L\ln(L)$ in 1D and 3D, 
respectively. For OBCs, the above KZ result (which is insensitive to boundary 
conditions) differs by $\ln(L)$ from the correct form in all dimensions.

Our results demonstrate that a scaling collapse in the ordering
behavior of a many body system is not per se evidence of KZ
scaling. On the contrary, scaling can arise from the dynamical
behavior of the excitations as the system relaxes into its ordered
state. By an appropriate effective modeling of these excitations, it
is possible to infer the dynamical scaling form of the order
parameter. A tell tale sign of the difference between KZ-driven and
defect-driven scaling may be observable when we compare the behavior
of open and closed boundary conditions, as we illustrate using examples in 
$d=1,2,3$. 

Finally, our work also provides analytical estimates (and
corresponding numerical verification) of the time scales relevant for
the onset of topological order as $T \to 0$ (following linear ramps in
temperature).  We remark that these time scales are indeed the ones
required to prepare the toric code in 2D and 3D in a topologically
ordered ground state devoid of any excitations. Even though the toric
code is but a toy model for topological quantum computing, modeling
of excitations in a manner similar to the one presented in our work
may be relevant to preparing quantum topological states in a potential
experimental setting for quantum information processing.


\begin{acknowledgments}

The authors would like to thank Paul Krapivsky for helpful discussions. 
This work was supported in part by the NSF under Grants No.~DMR-1410126 and No.~DMR-1710170 
(N.X. and A.W.S.) and by DOE Grant DE-FG02- 06ER46316 (C. Chamon), as well as by EPSRC Grant No.\ EP/K028960/1 
and EPSRC Grant No.\ EP/M007065/1 (C. Castelnovo). R.G.M was partially supported by the Natural Sciences and 
Engineering Research Council of Canada (NSERC), the Canada Research Chair program, and the Perimeter Institute 
for Theoretical Physics (supported by the Government of Canada through Industry Canada and by the Province 
of Ontario through the Ministry of Research \& Innovation). He also thanks Boston University's Condensed
Matter Theory Visitors Program for support. The computations were carried out on Boston University's 
Shared Computing Cluster. 
\end{acknowledgments}

\null


\vskip-5mm

\begin{thebibliography}{00}
\bibitem{wen_all} 
X.-G. Wen, Int. J. Mod. Phys. B 4, 239 (1990); Adv. Phys. {\bf 44}, 405(1995) ; Phys. Rev. B {\bf 65}, 165113 (2002). 

\bibitem{haldane85}
F. D. M. Haldane and E. H. Rezayi, Phys. Rev. B {\bf 31}, 2529 (1985).

\bibitem{wen90}
X.-G. Wen and Q. Niu, Phys. Rev. B {\bf 41}, 9377 (1990). 

\bibitem{eric02}
E. Dennis, A. Kitaev, A. Landahl, and John Preskill, J. Math. Phys. {\bf 43}, 4452 (2002).

\bibitem{nayak08}
 C. Nayak, S. H. Simon, A. Stern, M. Freedman and S. Das Sarma Rev. Mod. Phys. {\bf 80} 1083 (2008).

\bibitem{kitaev06}
A. Kitaev, Ann. Phys. {\bf 321}, 2 (2006). Ann. Phys.  N.Y.  303, 2  2003 

\bibitem{wegner71}
F. J. Wegner, J. Math. Phys. {\bf 12}, 2259 (1971).

\bibitem{john79}
J. B. Kogut Rev. Mod. Phys. {\bf 51}.659 (1979).

\bibitem{wansleben85}
S. Wansleben, J. Phys. A. Gen. {\bf 18}, L211 (1985).

\bibitem{claudio07}
C. Castelnovo and C. Chamon,  Phys. Rev. B {\bf 76}, 174416 (2007). 

\bibitem{macdonald11}
A. J. Macdonald, P. C. W. Holdsworth, and R. G. Melko, J. Phys.: Condens. Matter {\bf 23} 164208 (2011).

\bibitem{freeman14}
C. D. Freeman, C. M. Herdman, D. J. Gorman, K. B. Whaley, Phys. Rev. B {\bf 90}, 134302 (2014).

\bibitem{kibble76}
T. W. B. Kibble, J. Phys. A: Math. Gen. {\bf 9}, 1387 (1976).

\bibitem{zurek85}
 W. H. Zurek, Nature (London) {\bf 317}, 505 (1985).

\bibitem{zurek96}
W. H. Zurek, Phys. Rep. {\bf 276}, 177 (1996).

\bibitem{anatoli05}
A. Polkovnikov, Phys. Rev. B {\bf 72}, 161201(R) (2005).

\bibitem{grandi10}
C. De Grandi, V. Gritsev, and A. Polkovnikov, Phys. Rev. B {\bf 81}, 012303 (2010).

\bibitem{biroli10}
G. Biroli, L. F. Cugliandolo, and A. Sicilia, Phys. Rev. E {\bf 81}, 050101(R) (2010).

\bibitem{jelic11}
A. Jelic and L. F. Cugliandolo, J. Stat. Mech. {\bf 2011}, P02032 (2011).

\bibitem{hamp15}
J. Hamp, A. Chandran, R. Moessner, and C. Castelnovo, Phys. Rev. B {\bf 92}, 075142 (2015). 

\bibitem{anatoli11}
A. Polkovnikov, K. Sengupta, A. Silva, and M. Vengalattor, Rev. Mod. Phys. {\bf 83} 863 (2011).

\bibitem{ricateau17}
H. Ricateau, L. F. Cugliandolo, and M. Picco, J. Stat. Mech. (2018) 013201.

\bibitem{grandi11}
C. De Grandi, A. Polkovnikov, and A. W. Sandvik, Phys. Rev. B  {\bf 84}, 224303 (2011).

\bibitem{chandran12}
A. Chandran, A. Erez, S. S. Gubser, and S. L. Sondhi, Phys. Rev. B {\bf 86}, 064304 (2012).



\bibitem{zhong05}
F. Zhong and Z. Xu, Phys. Rev. B {\bf 71}, 132402 (2005).

\bibitem{liu14}
C.-W. Liu, A. Polkovnikov, and A. W. Sandvik, Phys. Rev. B {\bf 89}, 054307 (2014).

\bibitem{lamporesi13}
G. Lamporesi, S. Donadello, S. Serafini, F. Dalfovo, and G. Ferrari, Nature Phys. {\bf 9}, 656 (2013).

\bibitem{clark16}
L. W. Clark, L. Feng, C. Chin, Science {\bf 354}, 606 (2016).

\bibitem{liu15}
C.-W. Liu, A. Polkovnikov, A. W. Sandvik, and A. P. Young,  Phys. Rev. E {\bf 92}, 022128 (2015).

\bibitem{liu15a}
C.-W. Liu, A. Polkovnikov, and A. W. Sandvik, Phys. Rev. Lett. {\bf 114}, 147203 (2015).

\bibitem{rubin17}
S. J. Rubin, N. Xu, and A. W. Sandvik, Phys. Rev. E {\bf 95}, 052133 (2017).

\bibitem{xu17}
N. Xu, K.-H. Wu, S. J. Rubin, Y.-J. Kao, A. W. Sandvik, Phys. Rev. E {\bf 96}, 052102 (2017).

\bibitem{ben-av90}
R. Ben-Av, D. Kandel, E. Katznelson, P.G. Lauwer, and S. Solomon, J. Stat. Phys, Vol. {\bf 58}, 125 (1990).

\bibitem{savary17}
L. Savary and L. Balents, Rep. Prog. Phys. {\bf 80}, 016502 (2017).

\bibitem{claudio08}
C. Castelnovo and C. Chamon, Phys. Rev. B {\bf 78},155120 (2008).

\bibitem{showk14}
S. El-Show, M. F. Paulos, D. Poland, S. Rychkov, D. Simmons-Duffin, and A. Vichi,
J. Stat. Phys. {\bf 157}, 869 (2014). 


\bibitem{wansleben87}
S. Wansleben and D. P. Landau, Journal of Applied Physics {\bf 61}, 3968 (1987).

\bibitem{wang95}
F. Wang, N. Hatano, and M. Suzuki,  J. Phys. A: Math. Gen. {\bf 28}, 4543 (1995).

\bibitem{condamin05}
S. Condamin, O. Benichou, and M. Moreau, Phys. Rev. Lett. {\bf 95}, 260601 (2005). 

\bibitem{paul10}
P. L. Krapivsky, J. Stat. Mech. {\bf 2010}, P2014 (2010).

\bibitem{sidbook}
S. Redner, {\it A Guide to First-Passage Processes}, Cambridge University Press (Cambridge, 2001).

\end{thebibliography}
\end{document}